\documentclass[a4paper,11pt]{article}
\usepackage{times}
\usepackage{amsmath,amssymb, graphicx,epsfig,here}
\usepackage{xfrac}%for nicer fraction, like \sfrac{}{}
 \usepackage[colorlinks=true]{hyperref}
\hypersetup{urlcolor=blue, citecolor=red}
\providecommand{\keywords}[1]
{
  \small
  \textbf{\textit{Keywords---}} #1
}

\begin{document}

\title{Performance analysis of Volna-OP2 -- massively parallel code for tsunami modelling}
    \author{Daniel Giles$^{1}$, Eugene Kashdan$^{1}$, Dimitra M. Salmanidou$^{3}$,
    Serge Guillas$^{3}$, and Fr\'ed\'eric Dias$^{1,2}$\\
    {\footnotesize $^{1}$ School of Mathematics and Statistics, University College Dublin, Dublin, Ireland} \\
	{\footnotesize $^{2}$ Earth Institute, University College Dublin, Dublin, Ireland}\\
	{\footnotesize $^{3}$ Department of Statistical Science, University College London, London, UK}}
	\date{}

	\maketitle
	\abstract{The software package Volna-OP2 is a robust and efficient code capable of simulating the complete life cycle of a tsunami whilst harnessing the latest High Performance Computing (HPC) architectures. In this paper, a comprehensive error analysis and scalability study of the GPU version of the code is presented. A novel decomposition of the numerical errors into the dispersion and dissipation components is explored. Most tsunami codes exhibit amplitude smearing and/or phase lagging/leading, so the decomposition shown here is a new approach and novel tool for explaining these occurrences. It is the first time that the errors of a tsunami code have been assessed in this manner.

	To date, Volna-OP2 has been widely used by the tsunami modelling community. In particular its computational efficiency has allowed various sensitivity analyses and uncertainty quantification studies. Due to the number of simulations required, there is always a trade-off between accuracy and runtime when carrying out these statistical studies. The analysis presented in this paper will guide the user towards an acceptable level of accuracy within a given runtime.  \\
	}


	\section{Introduction}
The software package Volna-OP2 is used for the simulations of tsunami modelling by a number of research groups around the world since its first introduction in 2011 \cite{dutykh_Volna_2011}. The driving force for developing the code was a need within the tsunami research community for a solver which was applicable for analysis of realistic tsunami events and aimed to aid operational tsunami research \cite{dutykh_Volna_2011,Reguly_Volna_2018}. The code solves the depth-averaged Nonlinear Shallow Water Equations (NSWE) in two horizontal dimensions $ (x, y) $ using modern numerical methods for solution of hyperbolic systems. Volna-OP2 can efficiently simulate the complete life of a tsunami from generation induced by bathymetry displacement, propagation and inundation onshore. It can be used for cases of a simplified bathymetry represented by a mathematical formula but also for the complex bathymetry and topography of the examined geographical region. Owing to the use of an unstructured triangular mesh, irregular bathymetric and topographic features can be efficiently captured and represented.

The first operational use of Volna-OP2 for a realistic scenario was for the modelling of sliding and tsunami generation in the St. Lawrence estuary in Canada \cite{Poncet_2010}. The code has been used to model varying tsunamigenic episodes \cite{dias_modelling_2014,gopinathan_2017}; in several cases it has been used in conjunction with statistical modelling to perform comprehensive sensitivity analysis tests and uncertainty quantification \cite{stefanakis_can_2014,beck_sequential_2016, sarri_cascadia_2018, salmanidou_statistical_2017, Liu_Guillas_2017}.

Both the originally developed and the newly parallelised version of Volna-OP2 have been carefully validated against well known benchmarks available to the tsunami community \cite{dutykh_Volna_2011,Reguly_Volna_2018}. However, in the present paper, we pay particular attention to the accuracy of the new GPU version of the code with special emphasis on dispersion and dissipation errors as well as its computational efficiency on the general purpose GPU cluster. Gaining a deeper understanding of the numerical errors present can inform the user towards more accurate real case simulations. The manuscript is organised as follows: in the next section (\ref{model}), we describe the mathematical model and numerical schemes implemented in the parallel version of the code. Section (\ref{thacker}) is dedicated to an analytic benchmark used to explore the accuracy of the code and it is followed by the analysis of the dispersion and dissipation errors in Section (\ref{Error_analysis}). Section (\ref{Real_Cases}) discusses application of the code to real cases and the scalability of its GPU implementation. The paper is wrapped up by concluding remarks and perspective developments in Section (\ref{Conclusions}).

	\section{Mathematical Model and Algorithms}\label{model}

\subsection{Nonlinear shallow water equations}
The shallow water theory is efficiently used to describe the physics of long waves like tsunamis, which are characterised by very large wavelengths in comparison to the depth of the basin over which they propagate. Neglecting dispersion the NSWE yield:

	\begin{align}\label{eq:NSLW}
    \frac{\partial H}{\partial t}+ \nabla\cdotp (H\vec{u}) &= 0, \\
    \frac{\partial(H\vec{u})}{\partial t}+ \nabla \cdotp (H\vec{u}\otimes\vec{u} +\frac{g}{2}H^2 \textbf{I})&= - gH\nabla h,
    \end{align}
    where $ H=(h + \eta) $ is the total water depth, described as the sum of the time-dependent bathymetry $ h(x,y,t) $ and the free surface elevation $\eta(x,y,t) $,  $ \vec{u}(u,v) $ is the fluid velocity in the $ x $ and $ y $ horizontal directions, \textbf{I} is the identity matrix and $ g $ is the acceleration due to gravity. Provided that $ H > 0 $ the system is strictly hyperbolic. In the wet/dry transition the system starts to become non-hyperbolic since  $ H = 0 $ in a dry region. To deal with that an algorithm that solves the shoreline Riemann problem developed by \cite{Brocchini_2001} is implemented in the code.

\subsection{Spatial discretisation}
A cell-centered Finite Volume (FV) numerical method is used for the spatial discretization in Volna-OP2 \cite{dutykh_Volna_2011}. Other tsunami codes based on the finite volume approach include Tsunami-HySEA \cite{HySEA_2017} and GeoClaw \cite{GeoClaw_2011}. The numerical flux implemented in a numerical algorithm has to ensure that some standard conservation and consistency properties are satisfied: the fluxes from adjacent control volumes sharing an interface exactly cancel when summed and the numerical flux with identical state arguments reduces to the true flux of the same state.
\begin{figure}[H]
	\begin{center}
		\includegraphics[scale=0.5, trim={0 8cm 0 8cm}, clip]{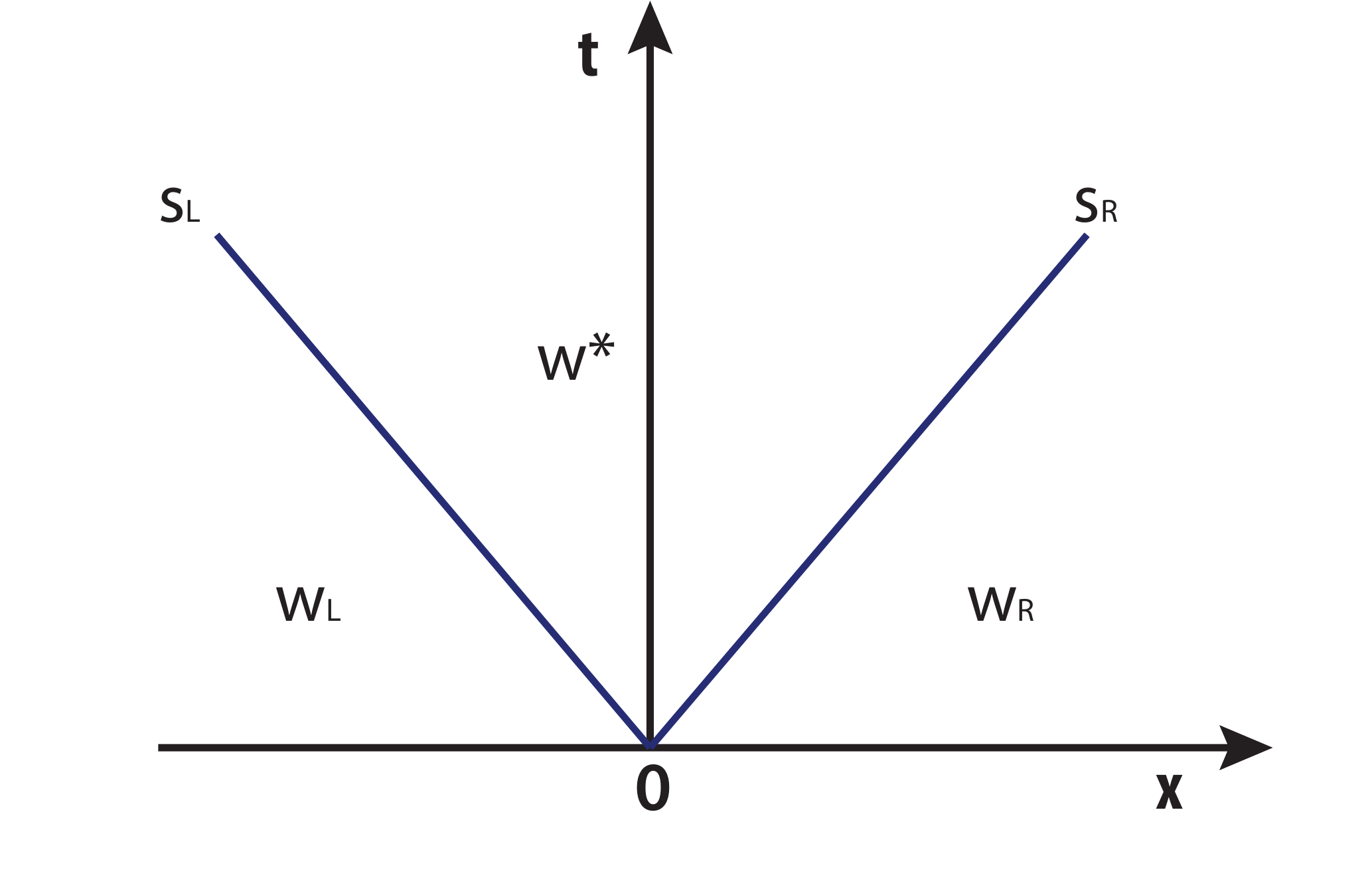}
		\caption{The approximate Riemann fan}\label{fig:fan}
	\end{center}
\end{figure}

In Volna-OP2 the Harten-Lax-van Leer (HLL) numerical flux was selected to ensure these conditions are met \cite{dutykh_Volna_2011}. The HLL approximate Riemann solver was proposed by Harten et al. in 1983 and assumes a two-wave configuration for the exact solution \cite{harten_hll_1983}. The wave speed chosen according to  \cite{Zhou_2002} yields a very robust approximate Riemann solver.
The Riemann solver models two waves that travel with speeds $ s_L $ and $ s_R $, the larger signal velocity is represented by $ s_R $ and the smaller by $ s_L $; three states are identified (Fig \ref{fig:fan}). The subscripts $ R $ and $ L $ are used to represent the right and left cell values respectively. The intermediate state is denoted by $ \vec{w}^* $, where $\vec{w}$ is a vector of the conserved variables $(H,Hu,Hv)$.
The numerical flux function of the scheme can be described by:
\begin{align*}
	\phi_{HLL}(\vec{w}_L,\vec{w}_R)&:=\left\{
	 \begin{array}{rcl}\vec{F}_L & \mbox{for} &  s_L\geq0, \\ \vec{F}^*  & \mbox{for} & s_L<0\leq{s_R}, \\\vec{F}_R  & \mbox{for} & s_R<0.
	 \end{array}
	 \right.
\end{align*}
where  $ \vec{w}_L,\vec{w}_R $ are the two interface states and $ \vec{F}_{L,^*,R} $ denotes the true flux at state $ \vec{w}_{L,^*,R} $ respectively. The right and left states are known. The intermediate state can be determined by applying the Rankine-Hugoniot conditions twice \cite{dutykh_Volna_2011}. It then derives that:
\begin{align*}
\vec{w}^*&=\frac{s_R \vec{w}_R-s_L \vec{w}_L-(\vec{F}_R-\vec{F}_L)}{s_R - s_L}
\\
\vec{F}^*&=\frac{s_R \vec{F}_L - s_L \vec{F}_R + s_L s_R(\vec{w}_R-\vec{w}_L)}{s_R-s_L}
\end{align*}
In Volna-OP2 the wave speeds are computed as:
\begin{align*}
 s_L&=min(u_{nL}-c_L, u^*_n-c^*),
 \\
 s_R&=max(u^*_n + c^*,u_{nR}+c_R)
\end{align*}
 where $ u_{nL}=\vec{u}_{L}\cdot \vec{n}_{LR}$, $ u_{nR}=\vec{u}_{R}\cdot \vec{n}_{LR} $ and $ u_n^* $ and $ c^* $ are equal to:
\begin{align*}
  u_n^*&=\frac{1}{2}(u_{nL}+u_{nR})+c_L-c_R,\\
  c^*&=\frac{1}{2}(c_L+c_R)-\frac{1}{4}(u_{nR}-u_{nL})
\end{align*}
where $  c_R=\sqrt{gH_R} $ and $ c_L=\sqrt{gH_L}  $ are the gravity wave speeds for the right and left state of the system respectively and $ \vec{n}_{LR} $ denotes the vector along the shared face between the right and left states.
The shortcoming of the HLL scheme is that it cannot resolve isolated contact discontinuities. It can thus become quite dissipative.

\subsection{Temporal discretisation}

A Strong Stability-Preserving (SSP) method is used in conjunction with a Runge-Kutta method for the temporal discretization in VOLNA. In the current version of the code the optimal second order two stage Runge-Kutta scheme SSP-RK(2,2) is used, with optimal Courant-Friedrichs-Lewy (CFL) condition equal to 1. The scheme is given as follows:
\begin{align*}
\vec{w}^{(1)}&=\vec{w}^{(n)}+\Delta t\mathcal{L}(\vec{w}^{(n)})\\
\vec{w}^{(n+1)}&=\frac{1}{2}\vec{w}^{(n)}+\frac{1}{2}\vec{w}^{(1)}+\frac{1}{2}\Delta t\mathcal{L}(\vec{w}^{(1)})
\end{align*}

Where $\mathcal{L}(\vec{w})$ is defined as the finite volume space discretization operator. The stability of the scheme is guaranteed if the CFL condition is satisfied. The Runge-Kutta scheme is very robust, especially in handling discontinuities. However, the scheme is both
dissipative and dispersive \cite{Howen_Sommeijer_1987}. Dissipativity causes a leak of energy from the system while dispersion leads to either phase lag or phase lead. A full explanation of these phenomena is given in Section (\ref{Error_analysis}).

\subsection{Second order extension}

The classical finite volume schemes are only first order accurate in space, which is insufficient for most modern computational simulations. Simulating a real tsunami case with a first order accurate scheme would require an unfeasible mesh resolution to obtain meaningful results. So in order to yield second order accuracy in space, a reconstruction technique is implemented. One must ensure that the scheme is total variation diminishing (TVD), i.e no artificial maxima and minima are introduced. Thus, within Volna-OP2 a MUSCL (Monotone Upstream-centered Scheme for Conservation Laws) scheme is implemented. The second order spatial accuracy is achieved by reconstructing the conserved variables on the cell interfaces. The reconstruction relies on calculating the gradient of a conserved variable over a cell and projecting the reconstructed value on the interface. Within Volna-OP2, a least squares gradient reconstruction and Barth-Jesperson limiter \cite{BarthJesperson1989} are implemented. The reconstructed values given below (\ref{eq_recon}) are then used in the numerical flux calculation:

\begin{equation}
\label{eq_recon}
\vec{w}(\vec{x_f})= \vec{w_{K}} + \alpha_{K} (\nabla \vec{w}) _K \cdot ( \vec{x_f} - \vec{x_0})
\end{equation}
where $\vec{w}$ is a vector of conserved variables, $\vec{w}(\vec{x_f})$ is the conserved variable evaluated at the interface, $\alpha_{K}$ is the cell specific conserved variable limiter, $ (\nabla \vec{w}) _K$ is the cell centred gradient and $ \vec{x_f} - \vec{x_0}$ is a vector pointing from the cell-centre to the face centre.
To avoid large gradients being calculated in the wet/dry region of the domain a threshold depth has been introduced to ensure that the code remains stable. When the depth goes below this threshold the scheme doesn't carry out a reconstruction. This threshold depth has been set to be $H_{\text{threshold}} =10^{-6}$m. However, it has been found that a conservative value can ensure greater stability, for example in Section (\ref{thacker}), $H_{\text{threshold}} =10^{-5}$m. For the real case (Section (\ref{Real_Cases})), $H_{\text{threshold}} =10^{-3}$m. At present these values are found through a trial and error approach but more research is required on the optimisation of this threshold depth.

\subsection{Boundary Conditions - Wall/Solid Boundary}
For a full explanation on the treatment of boundary conditions the reader is referred to \cite{dutykh_Volna_2011}. However, the case of a wall/solid boundary is given here. For all boundary conditions a ghost cell technique is used. This approach allows one to reconstruct the conserved variables on the boundary and thus preserve second order accuracy. Values of the conserved variables on the ghost cells are defined based on the type of boundary condition needed. In the following, cell $L$ is defined to be inside and cell $R$ (ghost cell) is outside of the domain. The boundary of the domain is the common edge between cell $L$ and cell $R$.

For a wall/solid boundary, $\vec{u} \cdot \vec{n} =0$, where $\vec{u}$ is the flow velocity and $\vec{n}$ is the normal vector to the boundary edge. To ensure that this is satisfied, the tangential ($\vec{u}^{\parallel}$) and normal ($\vec{u}^{\perp}$) velocities for the ghost cell are set to be equal and opposite to the those of the interior cell respectively:

\begin{align*}
    \vec{u}^{\perp}_{R} = -\vec{u}^{\perp}_{L} \\
    \vec{u}^{\parallel}_{R} = \vec{u}^{\parallel}_{L}
\end{align*}

\section{Benchmark Test with Analytical Solution}\label{thacker}

\begin{figure}[H]
\begin{center}
  \includegraphics[width=0.7\textwidth]{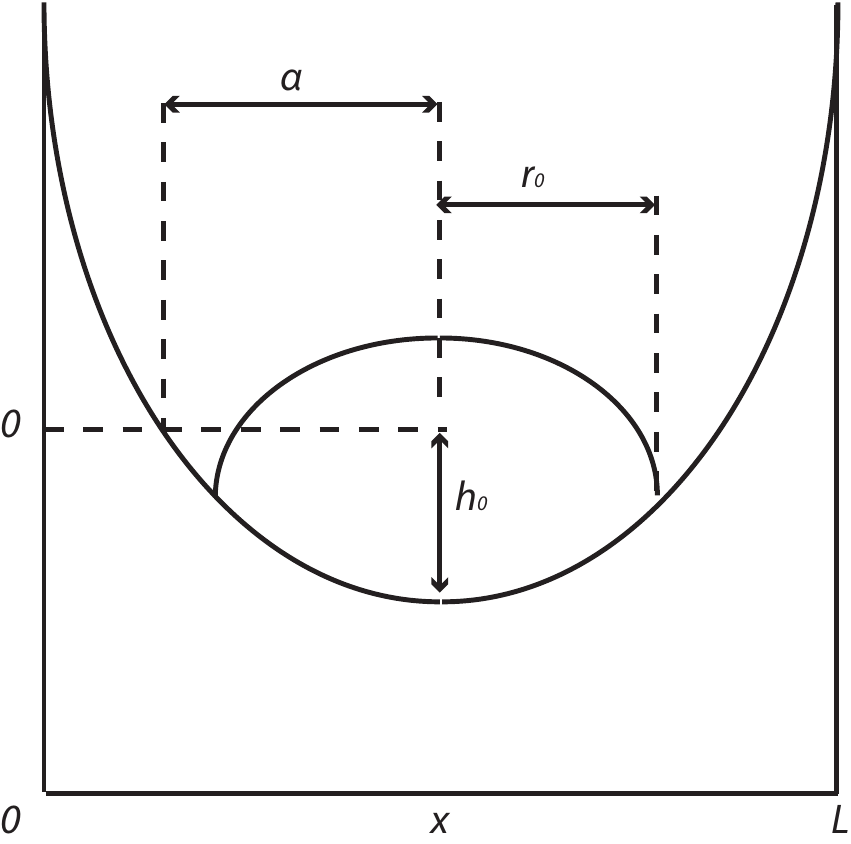}
  \caption{Geometry of the domain used for the analytic solution following \cite{Delestre_2013}}\label{fig:analytic}
\end{center}
\end{figure}

The two dimensional case of a radially symmetric paraboloid is implemented following the analytic solution initially proposed by Thacker \cite{Thacker_1981}. This solution is available in the SWASHES (Shallow-Water Analytic Solutions for Hydraulic and Environmental Studies) library \cite{Delestre_2013}. The major aim of SWASHES is to aid numerical modellers to validate shallow water equation solvers.
The oscillatory motion of the paraboloid is described by a periodic solution in which damping is assumed to be negligible. The morphology of the domain is a paraboloid of revolution given by:
\begin{align}
z(r)=-h_0\biggl(1-\frac{r^2}{\alpha^2} \biggr),
\end{align}
where $r=\sqrt{x^2+y^2}$ for each $(x, y)\in[-\frac{L}{2}, \frac{L}{2}] \times [-\frac{L}{2}, \frac{L}{2}]$, where $L$ is the length of the domain, $h_0$ is the water depth at the central point of the domain when the shoreline elevation is zero and $\alpha$ is the horizontal distance from the central point to the shoreline with zero elevation (Fig  \ref{fig:analytic}).

The free surface elevation $h(r, t)$ and the velocity components $u(x, y, t)$ and $v(x, y, t)$ are then given by:
\begin{align*}
h(r,t)&=h_0\biggl\{\frac{\sqrt{1-A^2}}{1-A\cos(\omega t)}-1-\frac{r^2}{\alpha^2}\biggl[\frac{1-A^2}{(1-A\cos(\omega t))^2}-1\biggr]\biggr\}-z(r),\notag\\
u(x,y,t) &= \frac{1}{1-A\cos(\omega t)}\biggl[\frac{\omega A\sin(\omega t)}{2}\biggl(x-\frac{L}{2} \biggr) \biggr],\\
v(x,y,t) &= \frac{1}{1-A\cos(\omega t)}\biggl[\frac{\omega A\sin(\omega t)}{2}\biggl(y-\frac{L}{2} \biggr) \biggr],\notag
\end{align*}
where  $\omega=\sqrt{8gh_0}/\alpha$ is the frequency, $r_0$ is the distance from the central point of the domain to the initial shoreline location and $A=(\alpha^2-r_0^2)/(\alpha^2+r_0^2)$. To model the solution we follow the values proposed in \cite{Delestre_2013}, where $\alpha=1$m, $r_0  = 0.8$m, $h_0 = 0.1$m, $L=4$m, and $T=3(2\pi/\omega)$.

\begin{figure}[H]
\begin{center}
  \includegraphics[width=\textwidth]{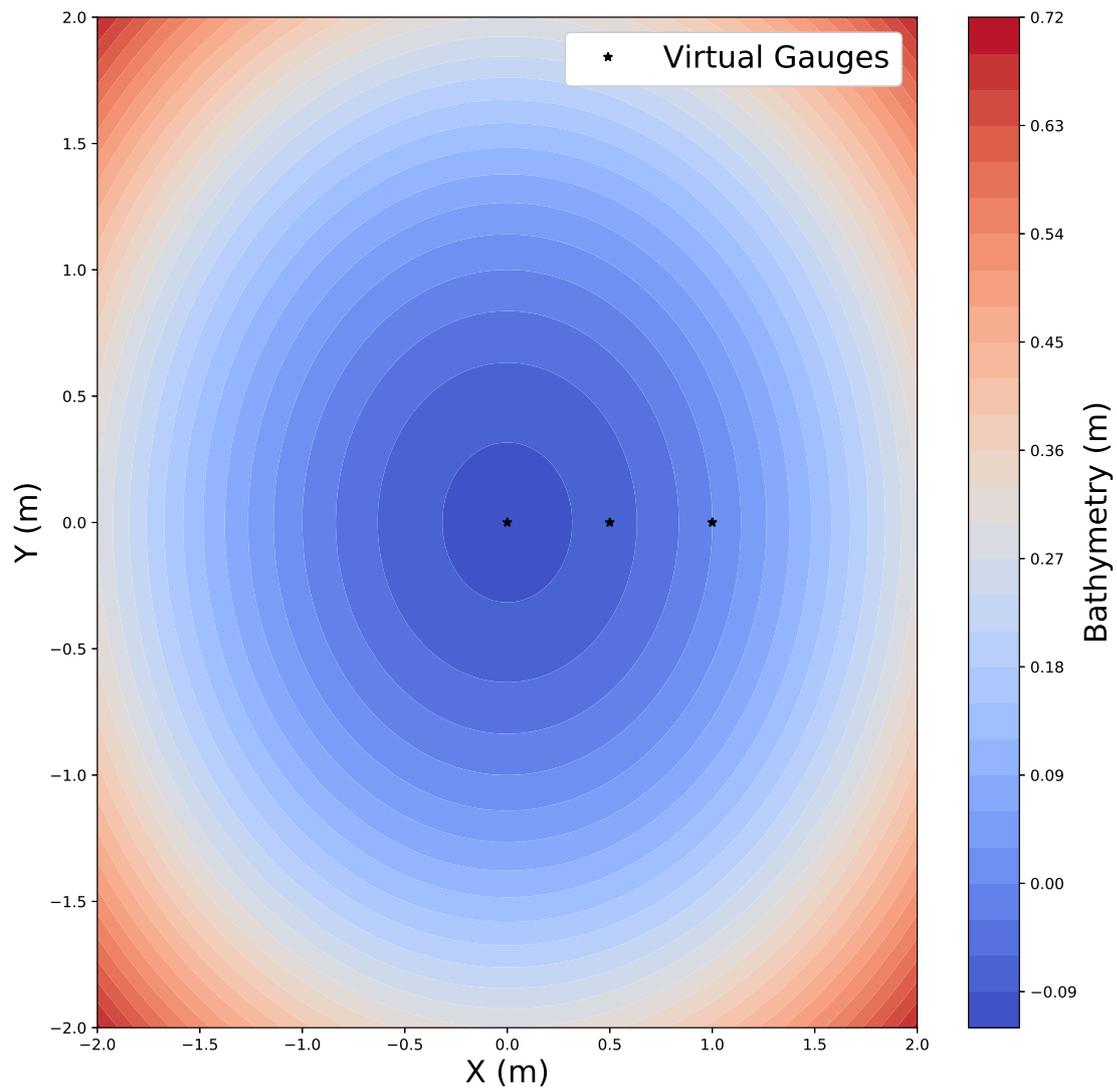}
\end{center}
\caption{Plot of the bathymetry (parabolic bowl) using the parameters outlined by \cite{Delestre_2013}. The colour coding matches the height of the bathymetry. The locations of the virtual gauges are marked by the black stars.}
\label{fig:thacker_domain}
\end{figure}

We record the free surface elevation in three positions $(x_1, y_1) = (0,0)$m (centre of the domain), $(x_2, y_2) = (0.5, 0)$m, and $(x_3, y_3) = (1, 0)$m (shoreline). Fig (\ref{fig:thacker_domain}) highlights a top-down view of the bathymetry and the locations of the wave gauges. In numerical simulations with Volna-OP2 we model the free surface elevation at the three positions with various spatial resolutions as a function of time up to $t_{fin}=10$s. An analytic solution at time $t=0$ is chosen as an initial condition and $\Delta t=0.45\Delta x$ for the simulations. This was chosen as it was found to be stable for all the mesh resolutions.

\begin{figure}[H]
\begin{center}
  \includegraphics[width=1\textwidth,height = 0.6\textwidth]{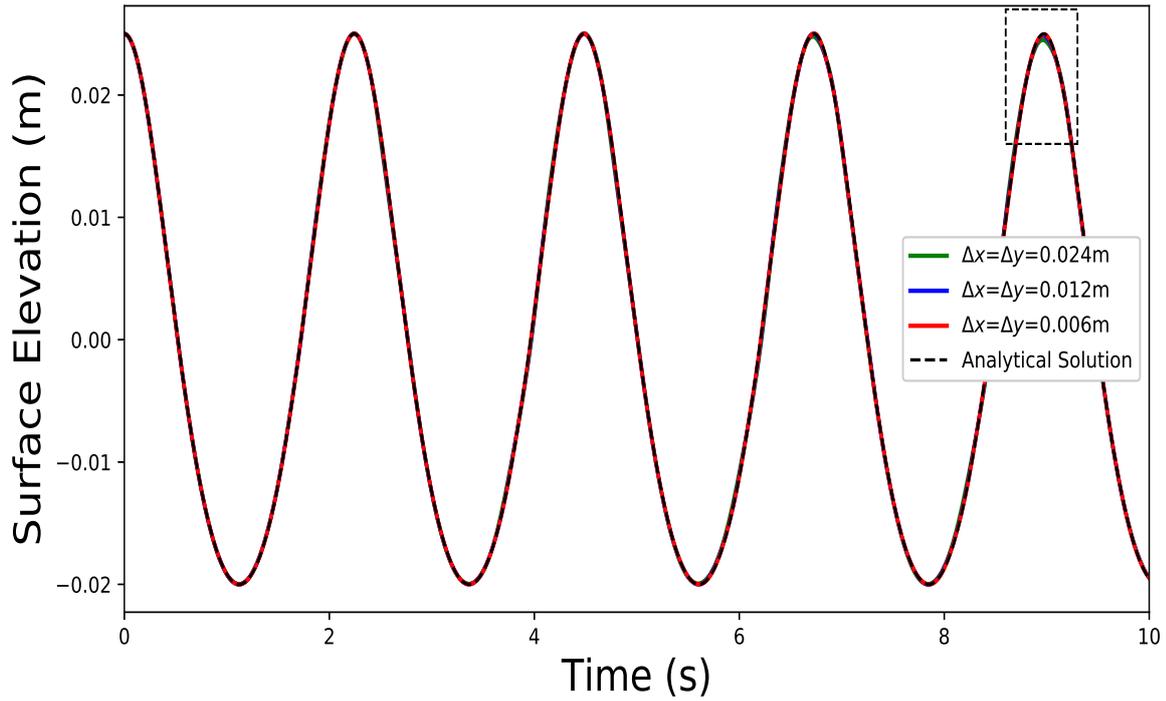}
\end{center}
\caption{Comparing numerical and analytic solutions in time at $(x_1, y_1) = (0, 0)$m (centre of the domain), where the analytic solution -- black dashed, $\Delta x=\Delta y= 0.024$m -- green,  $\Delta x=\Delta y= 0.012$m -- blue, and $\Delta x=\Delta y=0.006$m -- red. In the numerical simulations, $\Delta t=0.45\Delta x$. The dashed box is the boundaries of (Fig \ref{fig:center_zoom}).}
\label{fig:center}
\end{figure}

\begin{figure}[H]
\begin{center}
  \includegraphics[width=\textwidth]{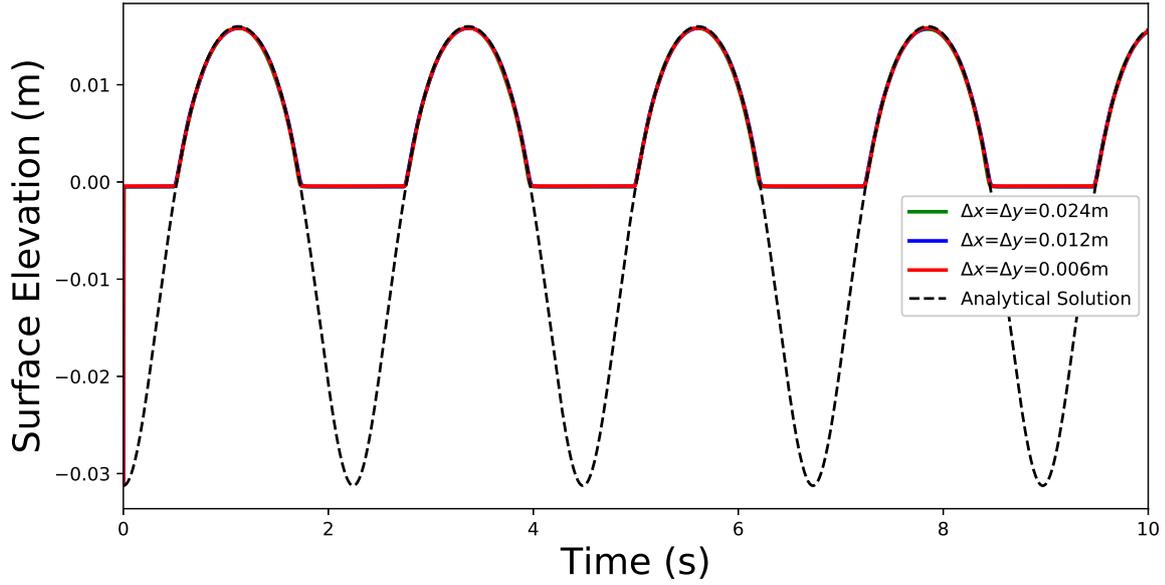}
\end{center}
\caption{Comparing numerical and analytic solutions in time at $(x_3, y_3) = (1, 0)$m (shoreline), where the analytic solution -- black dashed, $\Delta x=\Delta y= 0.024$m -- green,  $\Delta x=\Delta y= 0.012$m -- blue, and $\Delta x=\Delta y=0.006$m -- red. In the numerical simulations, $\Delta t=0.45\Delta x$. The shoreline forbids the numerical solutions to go below zero.}
\label{fig:shoreline}
\end{figure}

The results of the numerical simulations are shown in (Fig \ref{fig:center}) -- (Fig \ref{fig:shoreline}).

\begin{figure}[H]
\begin{center}
  \includegraphics[width=\textwidth]{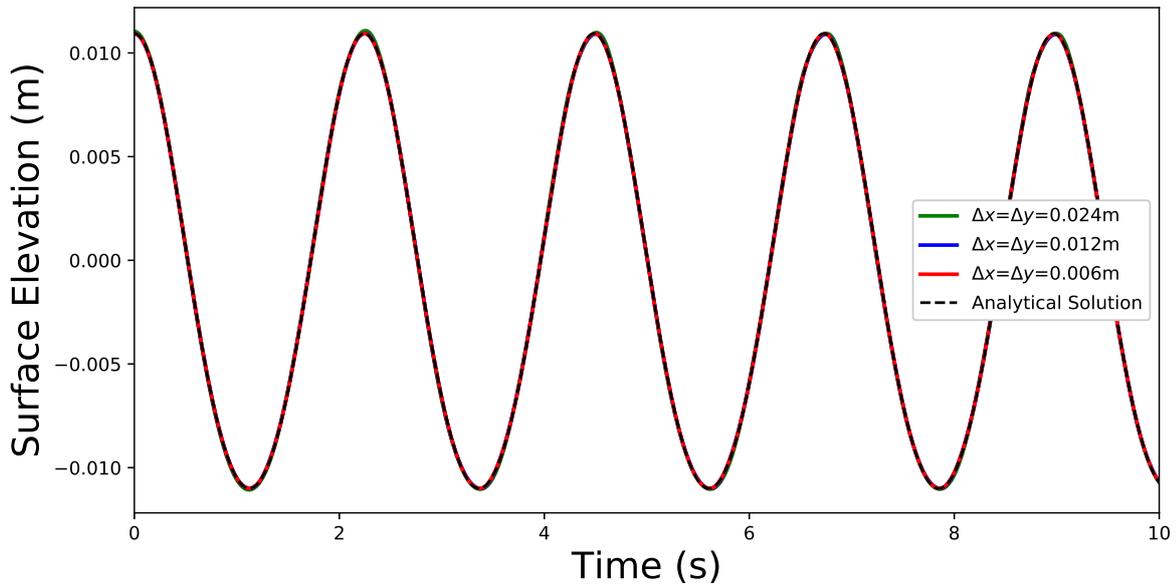}
\end{center}
\caption{Comparing numerical and analytic solutions in time at $(x_2, y_2) = (0.5, 0)$m, where the analytic solution -- black dashed, $\Delta x=\Delta y= 0.024$m -- green,  $\Delta x=\Delta y= 0.012$m -- blue, and $\Delta x=\Delta y=0.006$m -- red. In the numerical simulations, $\Delta t=0.45\Delta x$.}
\label{fig:middle}
\end{figure}

\begin{figure}[H]
\begin{center}
  \includegraphics[width=\textwidth]{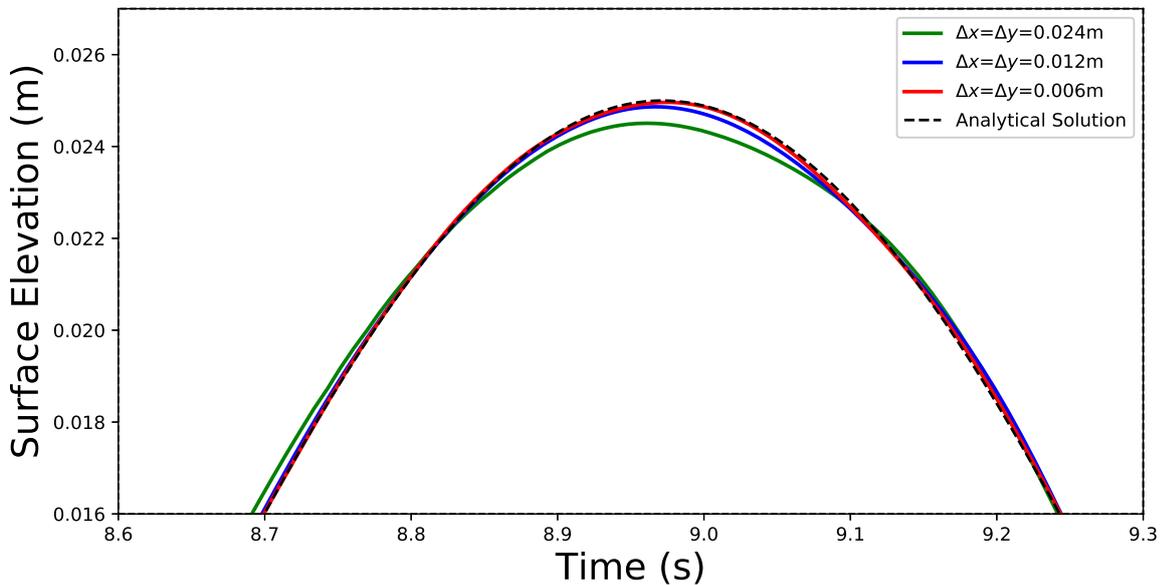}
\end{center}
\caption{Zoom in on the plot of the gauge at the centre of the domain $(x_1, y_1) = (0, 0)$m, one can observe the effect of the mesh size on the accuracy. }\label{fig:center_zoom}
\end{figure}

\begin{figure}[H]
\begin{center}
  \includegraphics[width=\textwidth]{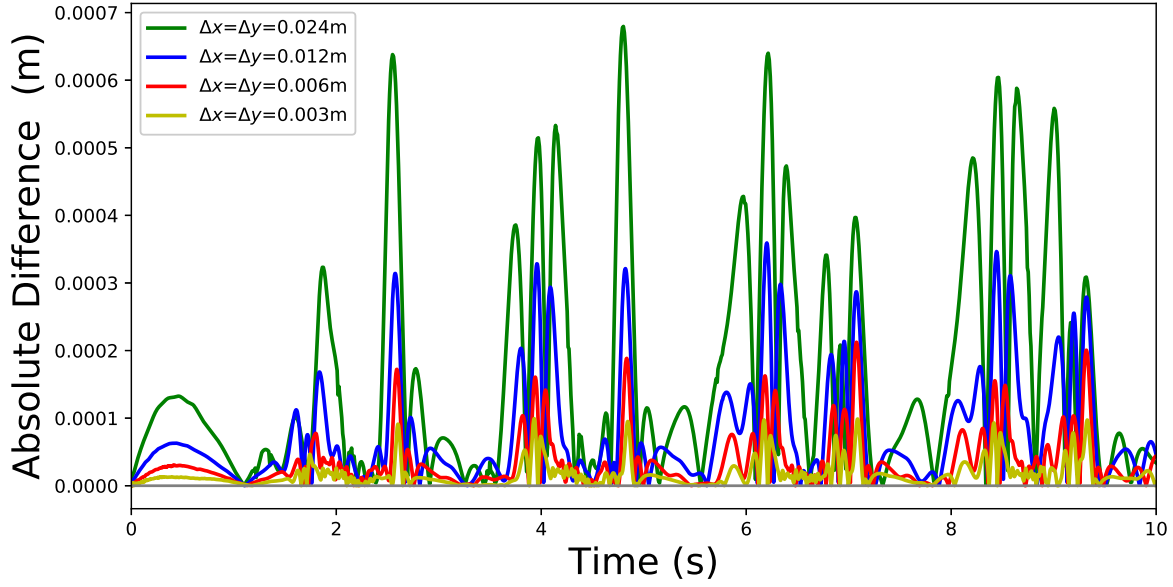}
  \caption{Absolute difference between the analytic and the numerical free surface elevation over time at the centre of the domain ($x=y=0$)m for spatial resolution: $\Delta x=\Delta y= 0.003$m -- yellow line, $\Delta x=\Delta y= 0.006$m -- red,  $\Delta x=\Delta y= 0.012$m  -- blue, $\Delta x=\Delta y= 0.024$m  -- green; $\Delta t=0.45\Delta x$.}
 \label{fig:abs_error}
\end{center}
\end{figure}

\begin{figure}[H]
\begin{center}
  \includegraphics[width=\textwidth]{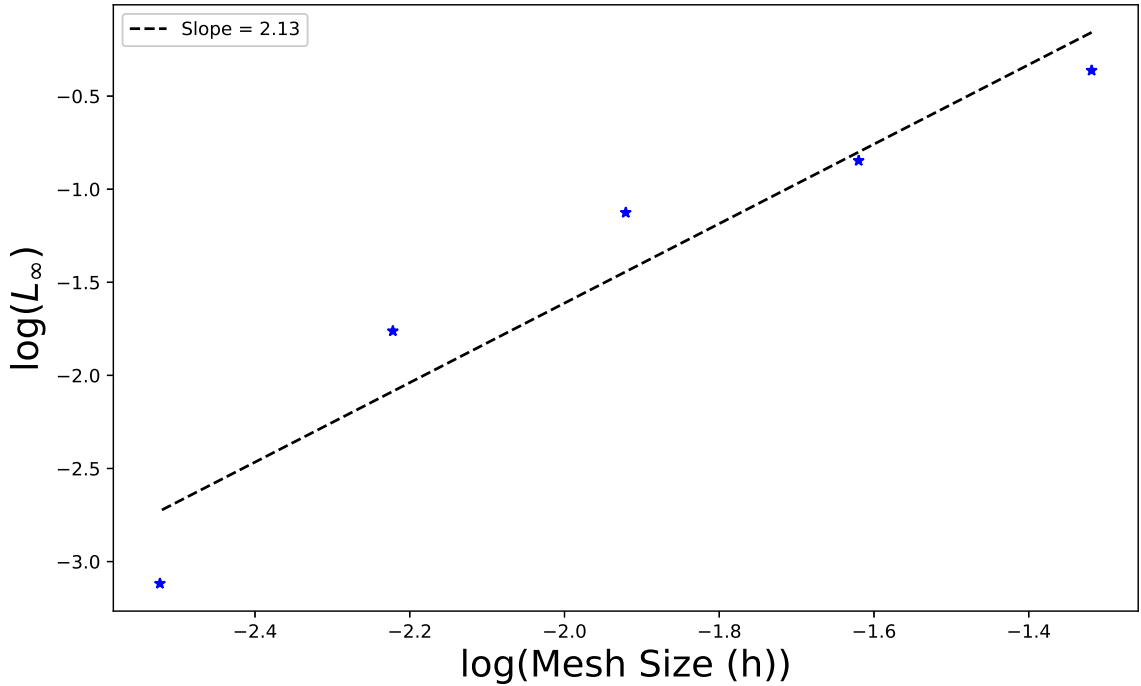}
  \caption{Convergence rate of the $L_{\infty}$ norm. Volna-OP2 with the MUSCL extension is $2^{nd}$ order accurate in space, i.e the error is $O(h^{2})$.}
  \label{fig:convergence}
\end{center}
\end{figure}

The finest mesh ($\Delta x=\Delta y=0.006$m) yields a representation closest to the surface elevation given by the analytic solution. The discrepancies between the meshes can only be seen by zooming in on the plots (Fig \ref{fig:center_zoom}). Focusing on the centre of the domain, we plot the absolute difference between the analytic and the numerical free surface elevation over time (Fig \ref{fig:abs_error}). As expected, the numerical error is always larger for the coarser meshes. In order to gain an idea on the numerical order of the scheme, a convergence study was carried out. The $L_{\infty}$ norm is calculated at 10s for various mesh resolutions (0.048 -- 0.003)m and then plotted versus the characteristic mesh size (Fig \ref{fig:convergence}).

The results shown in (Fig \ref{fig:abs_error}) and (Fig \ref{fig:convergence}) highlight that the scheme is second order accurate in space. However, the role of numerical dispersion and dissipation has not been explored and they should be accounted for when running long time tsunami simulations. We come back to this discussion in section (\ref{Error_analysis}).

\begin{figure}[H]
\begin{center}
  \includegraphics[width=\textwidth]{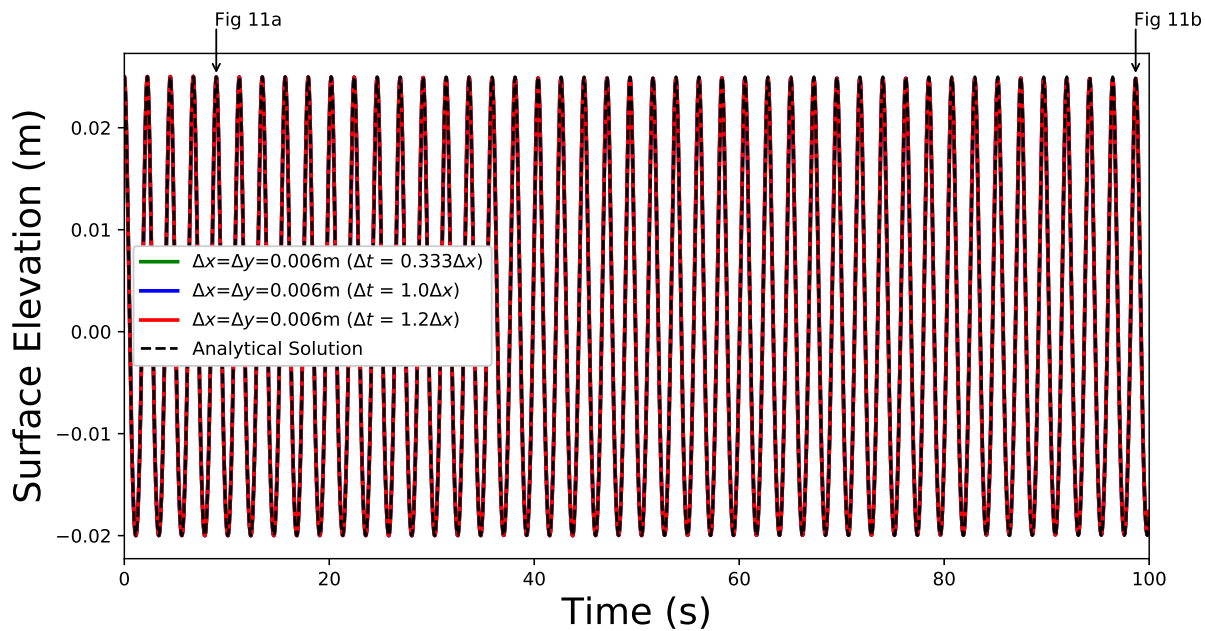}
\end{center}
\caption{Solution at the centre of the domain ($x=y=0$m) for various time steps with fixed spatial resolution $\Delta x=\Delta y= 0.006$m: the analytic solution -- black dash line, $\Delta t=0.333\Delta x$ -- green,  $\Delta t=\Delta x$ -- blue, and $\Delta t=1.2\Delta x$ -- red. Arrows point towards the areas highlighted in the subplots (Fig \ref{fig:timestep_zoom}a and Fig \ref{fig:timestep_zoom}b).}

\label{fig:timestep}
\end{figure}

\begin{figure}[H]
  \begin{minipage}[b]{\textwidth}
   \centering
   \includegraphics[width=0.6\textwidth]{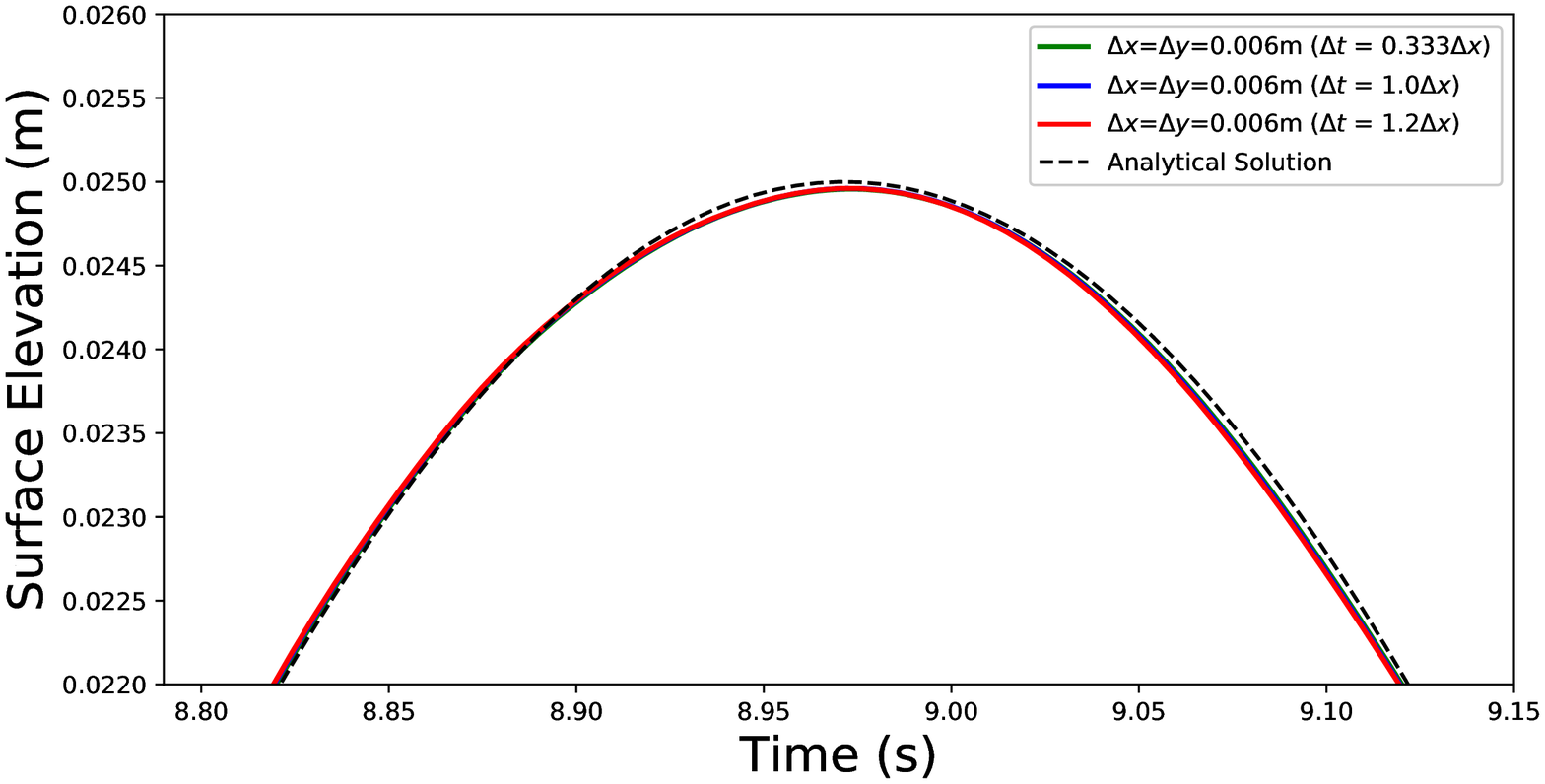}
   \vspace{4ex}
   \centering
   \includegraphics[width=0.6\textwidth]{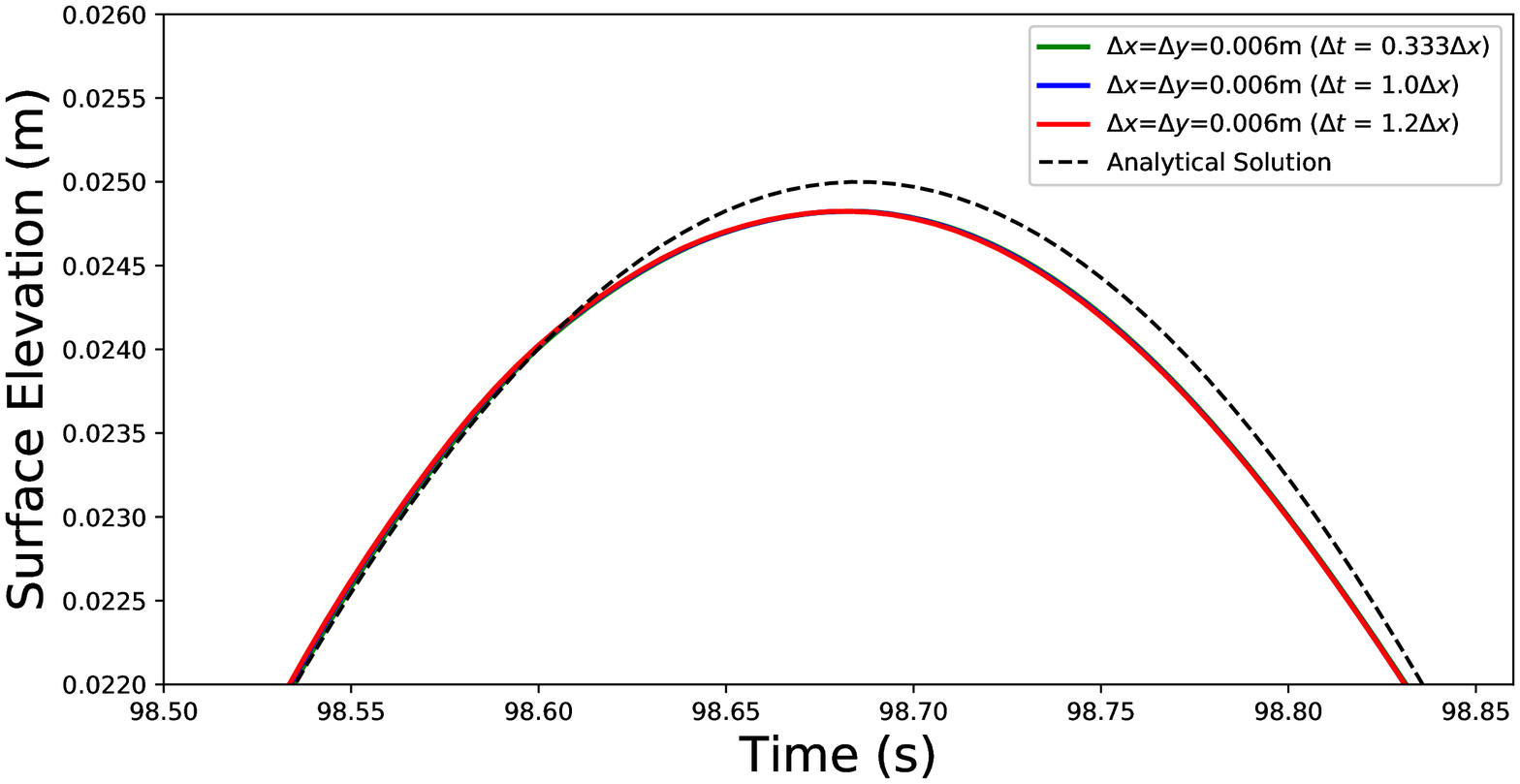}
   \vspace{4ex}
   \end{minipage}
   \caption{Zoom in on Fig \ref{fig:timestep} for two different time windows (Top (a): $t = 8.79-9.15$s and Bottom (b): $t = 98.5 - 98.86$s). The difference between the numerical solution with various time steps is not noticeable in each subplot. However, one can see a damping and phase shifting of the numerical signal when comparing the bottom (\ref{fig:timestep_zoom}b) and top subplot (\ref{fig:timestep_zoom}a).}
  \label{fig:timestep_zoom}
\end{figure}

To check the temporal discretization error we keep the mesh size constant at $\Delta x=\Delta y= 0.006$m and vary the time step by adjusting the constant connecting the spatial and temporal resolutions. We choose three values: $\Delta t=(0.333, 1, 1.2)\Delta x$ and run the test for a longer time $t_{fin}=100$s. The results of the simulations are shown in Figs \ref{fig:timestep} and \ref{fig:timestep_zoom}.

Changes in time-step within the stability limits while keeping the same spatial resolution almost do not affect the accuracy of the solution (Fig \ref{fig:timestep}). This is highlighted in Fig \ref{fig:timestep_zoom} where the plots of the numerical solution with various time steps overlap each other.  However, comparing the subplots of Fig \ref{fig:timestep_zoom} as time goes on, one observes a damping of the numerical signal and a phase shift. The Runge-Kutta scheme implemented in the code is dissipative and we could expect a leak of the energy from the system, thus we can expect the damping of the numerical signal. Reasons for the phase shift in the signal will be explained in section (\ref{Error_analysis}). Overall, the results from this section show that the space discretisation has a strong influence on the solution, while reducing the time step has no visible effect.

\section{Error analysis}\label{Error_analysis}

The  exact solution of the discretized equations satisfies a PDE which is generally different from the one to be solved. The original equation is replaced with the modified equation $Au^{n+1}=Bu^n$, or, in other words
\begin{align*}
\frac{\partial \omega}{\partial t}+\mathcal{L}\omega=0 ~~\text{becomes}~~
\frac{\partial \omega}{\partial t}+\mathcal{L}\omega=\sum^{\infty}_{p=1}\alpha_{2p}\frac{\partial^{2p}\omega}{\partial x^{2p}}+\sum^{\infty}_{p=1}\alpha_{2p+1}\frac{\partial^{2p+1}\omega}{\partial x^{2p+1}}.
\end{align*}
The even-order derivatives on the right-hand side produce
an amplitude error, or numerical dissipation. The odd-order derivatives on the right-hand side produce a wave-number-dependent phase error known as numerical dispersion. In the long time simulations, the numerical behaviour of the  scheme largely depends on the role played by the dispersive and dissipative effects also known as "wiggles" (phase errors) and "smearing" (amplitude errors) respectively. A negative dispersion coefficient corresponds to phase lagging (i. e. harmonics travel too slowly), while positive dispersion coefficients yield phase leading with spurious oscillations occurring ahead of the wave.

\begin{figure}[H]
\begin{center}
  \includegraphics[width=\textwidth]{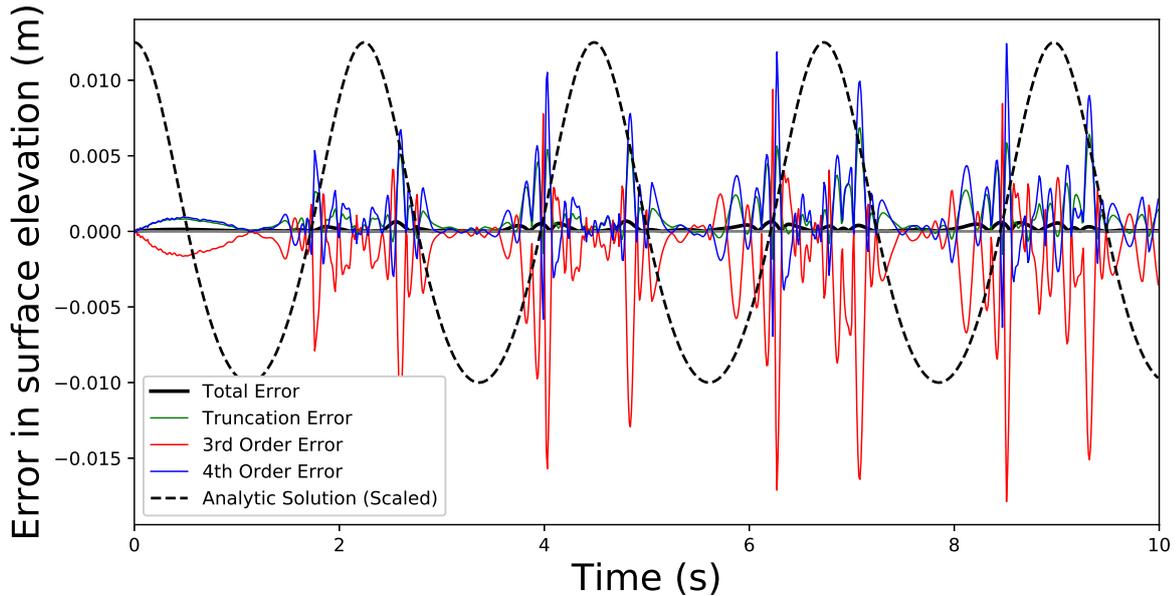}
\end{center}
\caption{Decomposition of the absolute error at the centre of the domain ($x=y=0$m) for a spatial grid with resolution of $\Delta x=\Delta y =0.024$m and CFL=0.45. The colours correspond to: the truncation error -- green,  the third order error -- red, the fourth order error -- blue, the total error -- black. The black dashed line is a scaled plot of the surface elevation at the centre of the domain over time.}
\label{fig:error24}
\end{figure}

\begin{figure}[H]
\begin{center}
  \includegraphics[width=\textwidth]{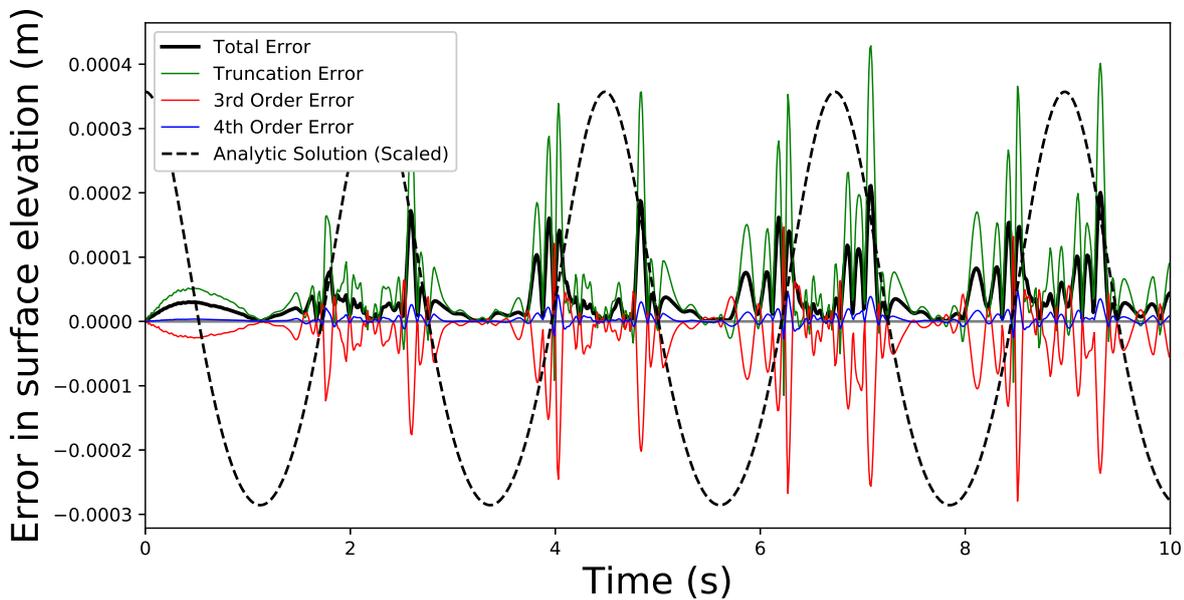}
\end{center}
  \caption{Decomposition of the absolute error at the centre of the domain ($x=y=0$m) for a spatial grid with resolution of $\Delta x=\Delta y =0.006$m and CFL=0.45. The colours correspond to: the truncation error -- green,  the third order error -- red, the fourth order error -- blue, the total error -- black. The black dashed line is a scaled plot of the surface elevation at the centre of the domain over time. }
  \label{fig:error06}
\end{figure}

According to the {\it Lax-Richtmyer Equivalence Theorem} \cite{Lax_Richtmyer_1956}, if a scheme has a truncation error of order $(p, q)$ and the scheme is stable, then the difference between the analytic solution and the numerical solution in an appropriate norm is of the
order $(\Delta t)^p + h^q$ for all finite time. It has been observed numerically (see Fig \ref{fig:convergence}) that the numerical solution is second order accurate. Taking into account that the time step is proportional to the spatial resolution that we call $h$, we can write that the total error is of the order $O(h^{2})$. To analyse the role played by the dissipation and dispersion errors, we rewrite the error as
\begin{align*}
E(t)=\sum_{p=2}^{\infty}C_{p-1}(t)h^p,
\end{align*}
where $C_{p-1}(t)$ incorporates all the constants included in the error formula. We choose the three leading terms of this expansion:
\begin{align*}
E(t)\approx C_1(t)h^2+C_2(t)h^3+C_3(t)h^4.
\end{align*}
The first term in this expansion corresponds to the truncation error (also the leading dissipation error). It is followed by the leading dispersion and the secondary dissipation error terms. We assume that the remaining terms are significantly smaller and can be neglected. Next we go back to the simulations with various spatial resolution discussed in Section (\ref{thacker}). For each of the three grids, we have the absolute error as a function of time. If we define the spatial resolution $h=0.024$m, two other grids have the resolution $h/2$ and $h/4$. The system of equations has the form:
\begin{align}
\sum_{p=2}^4C_{p-1}(t)\frac{h^{p}}{({2^{k-1}})^{p}}=E_k(t),~~k=1,2,3
\end{align}
and its solution is
\begin{align*}
C_1(t)&= \frac{E_1(t) - 24E_2(t) + 128E_3(t)}{3 h^2},\\
C_2(t)&= -\frac{2 [E_1(t) - 20E_2(t) + 64 E_3(t)]}{h^3},\\
C_3(t)&= \frac{8 [E_1(t) - 12E_2(t) + 32E_3(t)]}{3 h^4}
\end{align*}
The plots below show the composition of the total error for each of the grids (Fig \ref{fig:error24} -- Fig \ref{fig:error06}) , with a scaled surface elevation overlaid to give an idea on when the errors occur.

\begin{figure}[H]
\begin{center}
  \includegraphics[width=\textwidth]{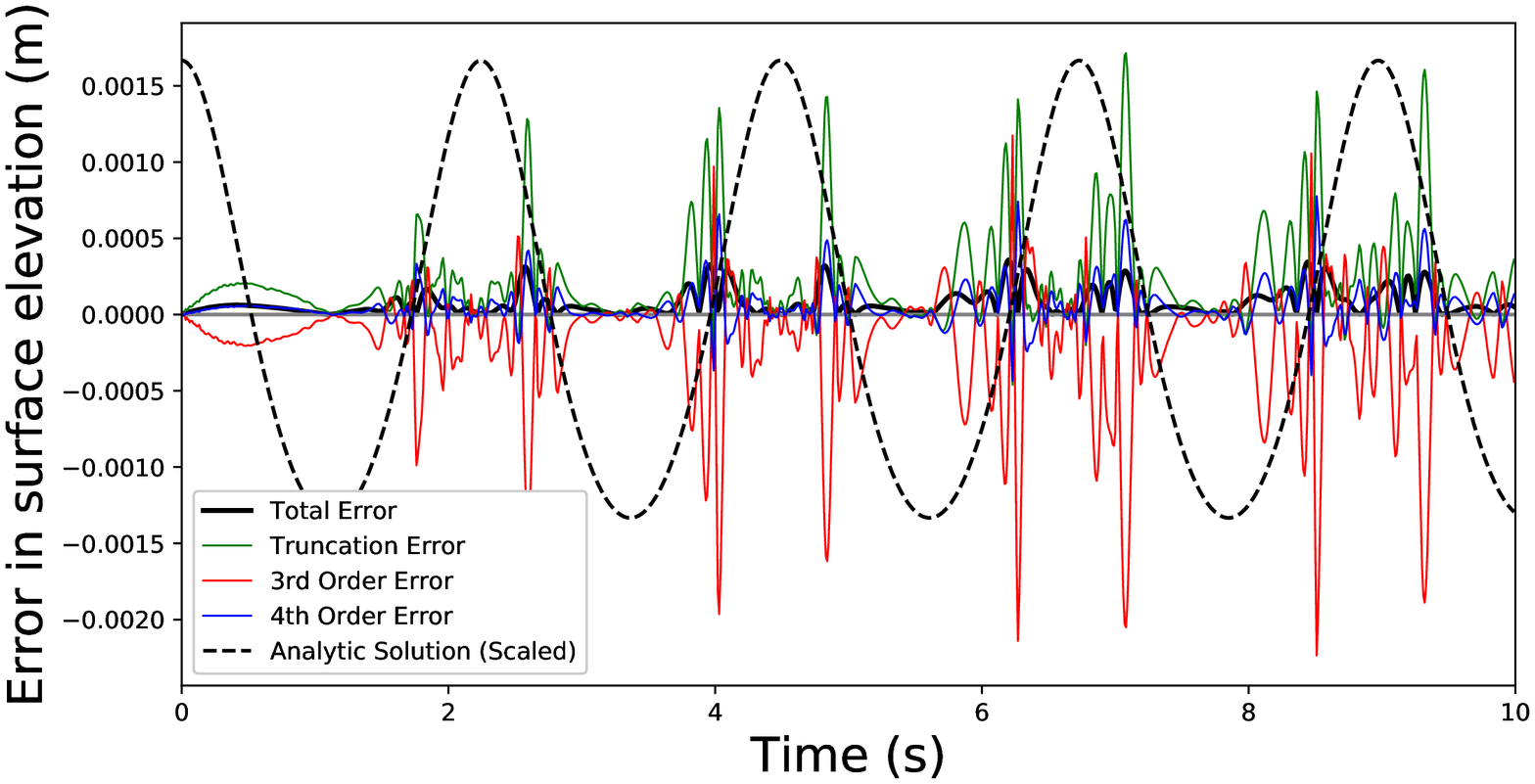}
\end{center}
\caption{Decomposition of the absolute error at the centre of the domain ($x=y=0$m) for a spatial grid with resolution of $\Delta x=\Delta y =0.012$m and CFL=0.45. The colours correspond to: the truncation error -- green,  the third order error -- red, the fourth order error -- blue, the total error -- black. The black dashed line is a scaled plot of the surface elevation at the centre of the domain over time.}
\label{fig:error12}
\end{figure}

In all the error decompositions, the components tend to cancel each other, which results in the total error being less than the individual components.  However, the leading dispersion error is a dominant component for all the total errors. The leading dispersion error exhibits large negative spikes, this points towards phase lagging as the surface elevation changes from negative to positive at the center of the domain. These large negative spikes in the leading dispersion error coincide with positive spikes in either the truncation or $4^{th}$ order dissipative errors. For finer grids the truncation error (leading dissipation error) plays a dominant role. For the larger grid the negative spikes are balanced by the $4^{th}$ order dissipative error (Fig \ref{fig:error24}).

This error analysis is important when considering real cases -- see section (\ref{Real_Cases}) -- as any error could be dominated by either the leading dispersion or dissipation terms. However, this numerical dispersion term could compensate for the fact that physical dispersion is neglected in the nonlinear shallow water equations, as shown by \cite{Burwell2007}.

\section{Real cases}\label{Real_Cases}

In this section we discuss an actual tsunami simulation done with Volna-OP2 running on the general purpose GPU cluster. The cluster consists of two NVIDIA\textsuperscript{\textregistered} Tesla\textsuperscript{\textregistered} V100 cards, with 5,120 CUDA cores, 16GB max memory size each.

\begin{figure}[H]
\begin{center}
  \includegraphics[width=1\textwidth]{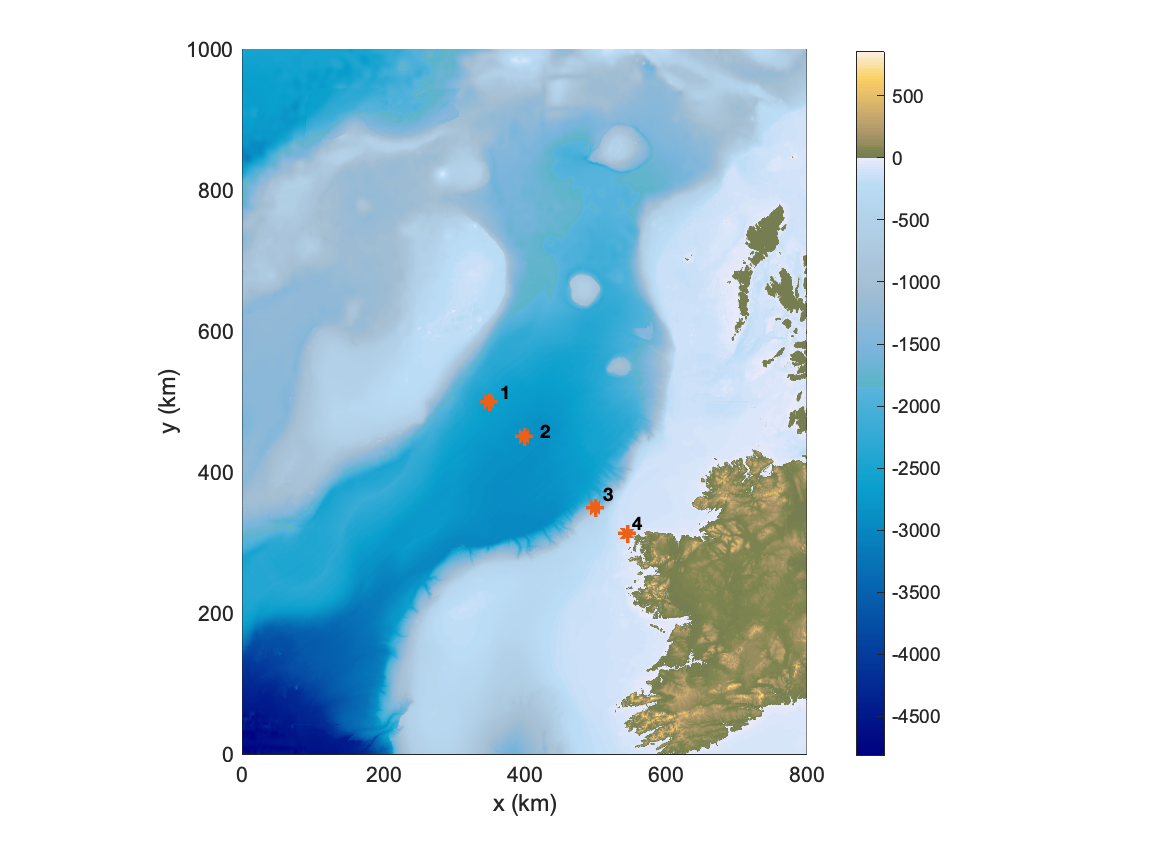}
  \caption{Computational domain used in the simulation and four gauges marked with the red dots.}
  \label{fig:gauges}
\end{center}
\end{figure}

\begin{figure}[H]
  \begin{minipage}[b]{0.5\textwidth}
   \centering
   \includegraphics[width=1\textwidth]{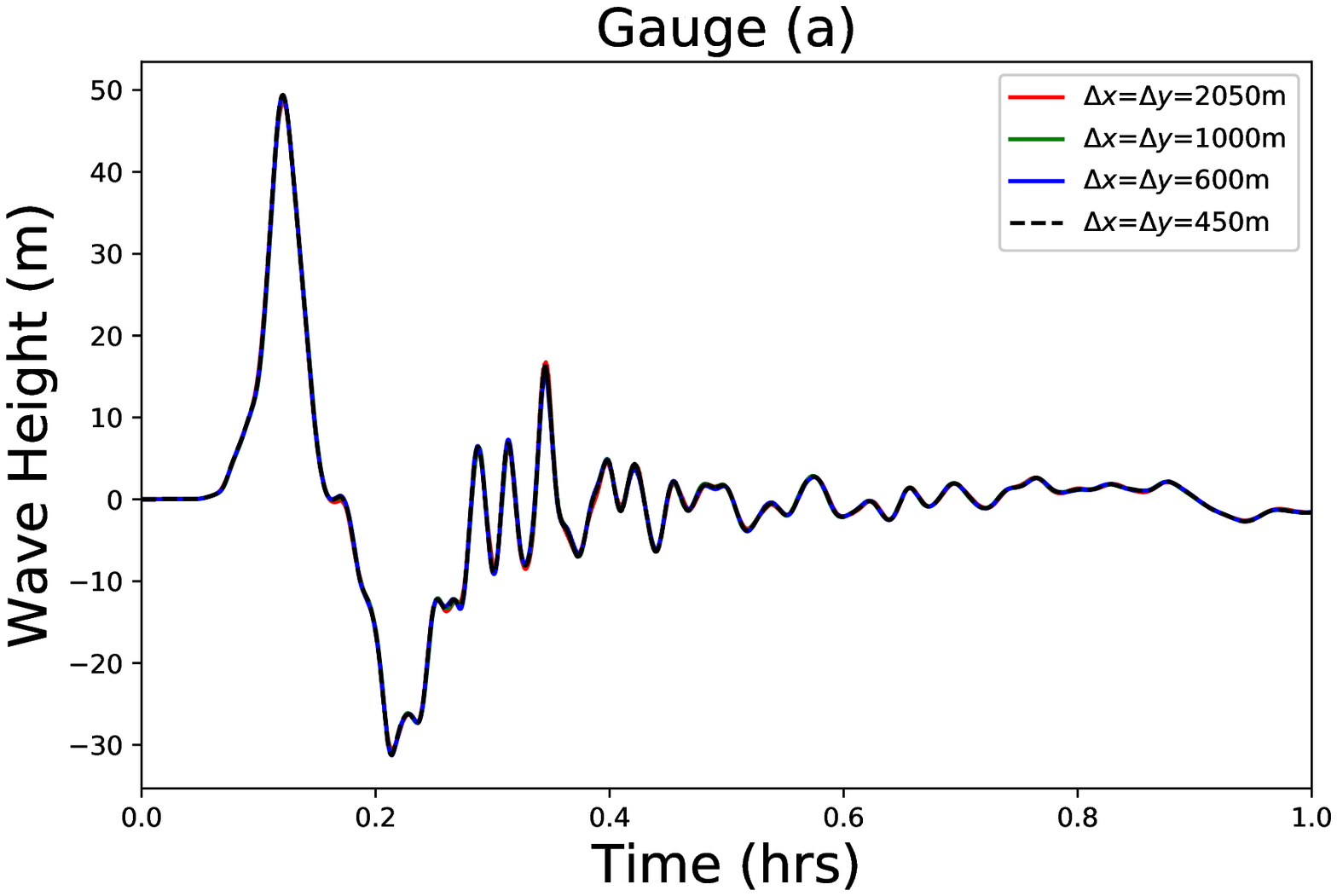}
   \vspace{4ex}
   \end{minipage}
   \begin{minipage}[b]{0.5\textwidth}
   \centering
   \includegraphics[width=1\textwidth]{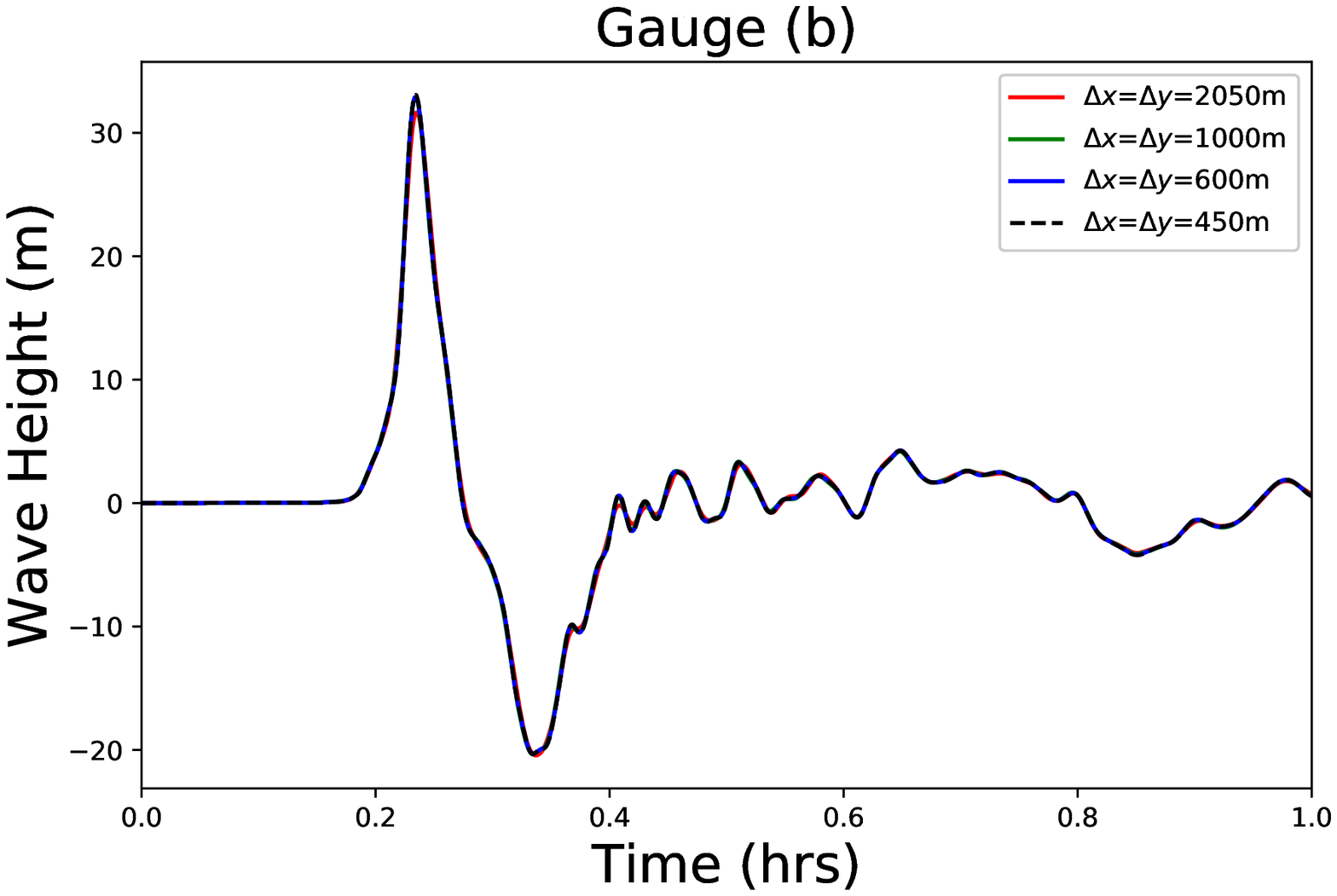}
   \vspace{4ex}
   \end{minipage}
  \begin{minipage}[b]{0.5\textwidth}
   \centering
   \includegraphics[width=1\textwidth]{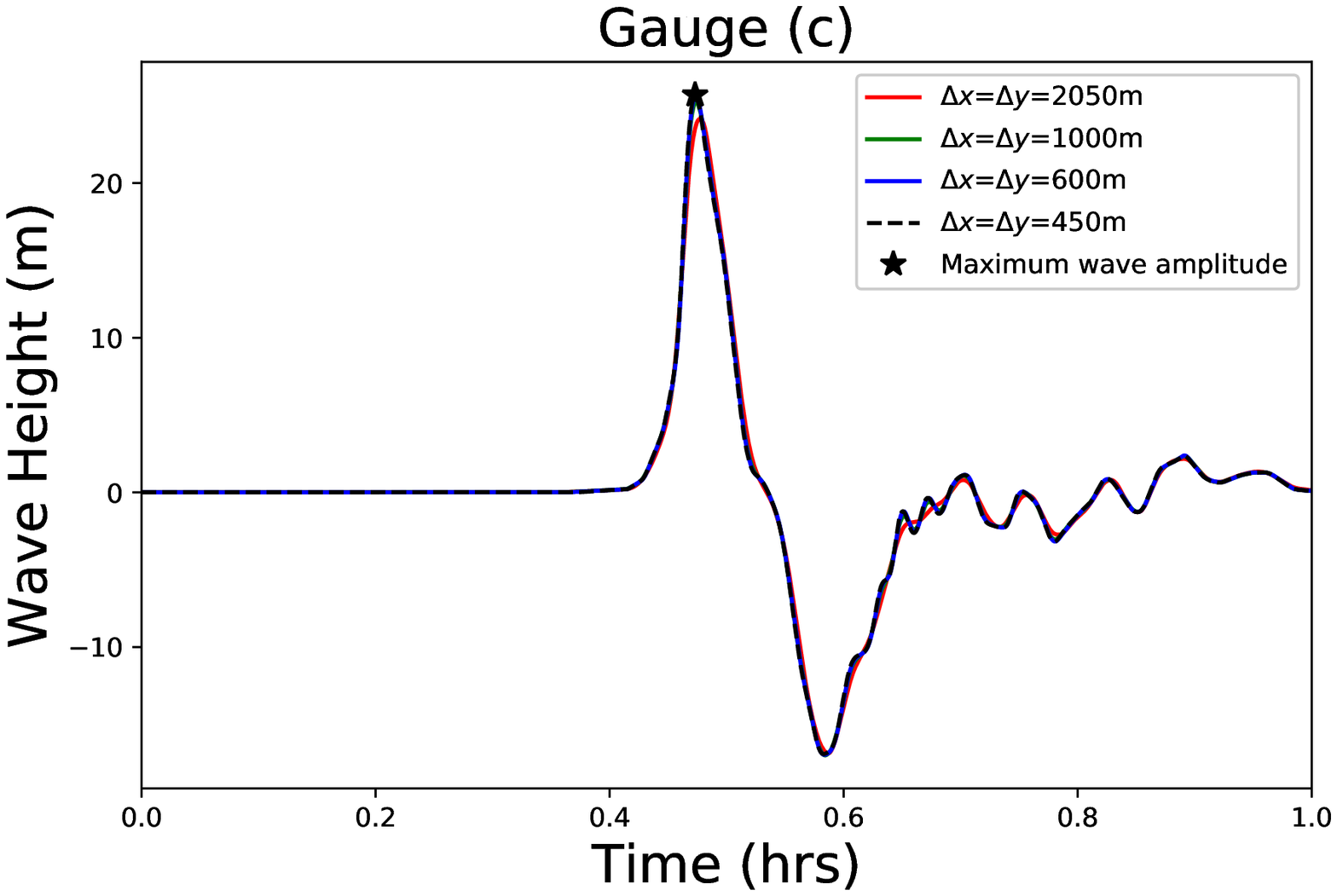}
   \vspace{4ex}
    \end{minipage}
   \begin{minipage}[b]{0.5\textwidth}
   \centering
   \includegraphics[width=1\textwidth]{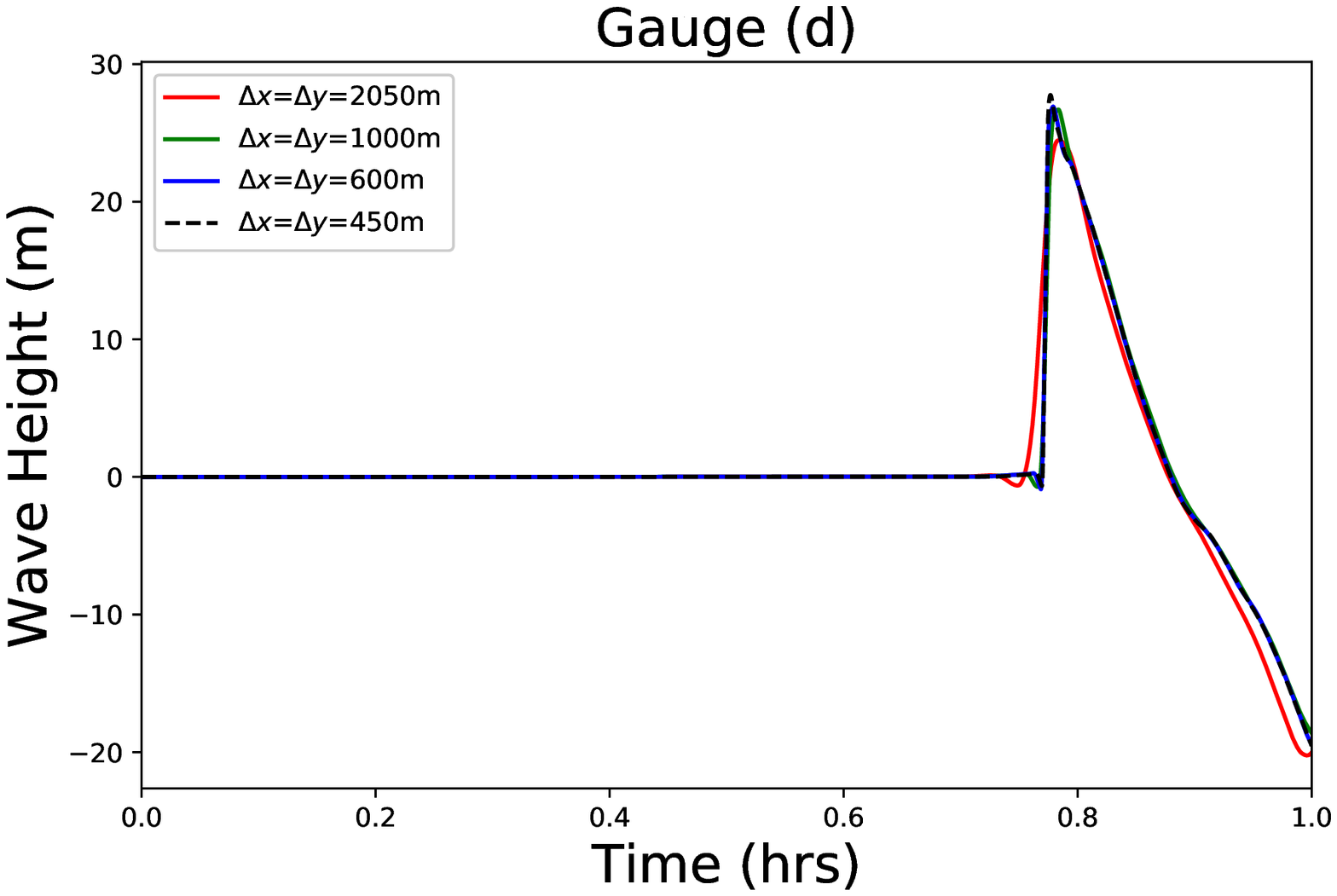}
   \vspace{4ex}
    \end{minipage}
  \caption{Results of the numerical simulations at four gauges from left to right on Fig \ref{fig:gauges}, sorted horizontally starting from the upper left corner. Each plot includes four tests with different spacial resolution: $\Delta x=\Delta y=$(2050m - red line, 1000m - green line, 600m - blue line, 450m - black dashed line) run for 1 hour with CFL=1.0. The gauges coordinates (from left to right, from top to bottom) are: a - (350km, 500km), b - (400km, 450km),  c - (500km, 350km), and d - (545.8km, 312.2km).}
  \label{fig:real_case}
\end{figure}

The domain size is $800 \times 1000$km and the physical simulation time is $1$ hour. The simulation corresponds to a hypothetical scenario of edge volume collapse (765 km$^{3}$) at the Rockall Bank Slide Complex. A geophysical study of the event taking into account volumetric, rheological and multiphase collapse considerations has been done under a different framework \cite{salmanidou_rheological_2018}. Convergence using a simple approach of material with visco-plastic rheology sliding in one go is demonstrated. For this study we have chosen four gauges marked by the red dots on Fig \ref{fig:gauges} to present the evolution of the tsunami as a function of time in Fig \ref{fig:real_case}.

\begin{figure}[H]
\begin{center}
  \includegraphics[width=1\textwidth]{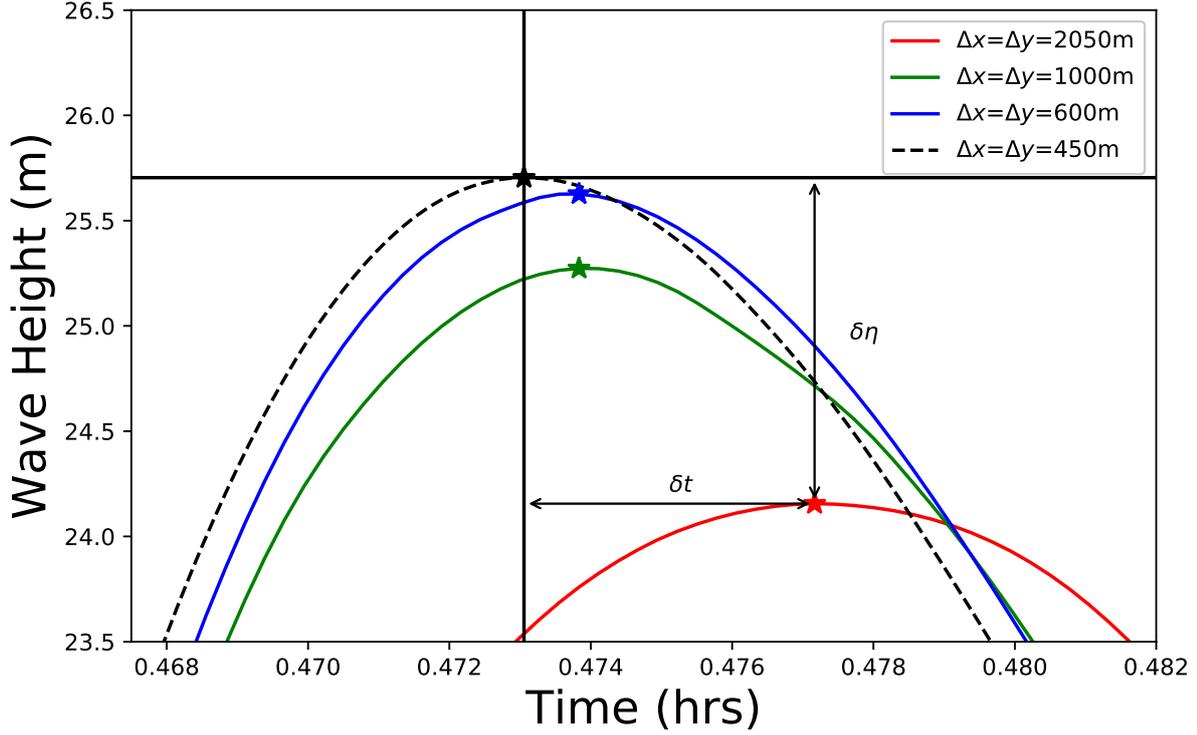}
  \caption{Zoom in on the maximum wave amplitudes of the initial tsunami wave at gauge 3 (Fig \ref{fig:real_case}c) simulated using the varying mesh resolutions}
  \label{fig:rbsc_max_amplitudes}
\end{center}
\end{figure}

Overall, the numerical simulations with varying spatial discretisation behave similarly. Notable differences can be seen at gauge 4 (Fig \ref{fig:real_case}d), where unresolved bathymetric features of the continental shelf play a role. To investigate the numerical differences we will focus on the output at gauge 3 (Fig \ref{fig:real_case}c). The reason for choosing this gauge is to minimise the effects of these unresolved bathymetric features as the bathymetry between this gauge location and the landslide source is relatively flat. The maximum wave amplitude of the initial tsunami wave at gauge 3 (Fig \ref{fig:real_case}c) has been highlighted above. As there is no true solution to compare with for this real case we will take the simulation results from the finest mesh $\Delta x=\Delta y = 450$m as the {\it ground truth}. Relative differences between this {\it ground truth} and the other simulations are presented in Fig \ref{fig:rbsc_max_amplitudes} and table \ref{tab:relative}. When comparing the signals (Fig \ref{fig:rbsc_max_amplitudes}), the coarser meshes exhibit phase lagging and/or damping of the signal, i.e. the maximum tsunami wave arrives later and its amplitude is diminished. This behaviour was highlighted in the previous error analysis -- Section \ref{Error_analysis} -- and is thus expected.

It should be noted that for cases which utilize non-uniform mesh resolutions the same error analysis findings will hold true. If the mesh is non-uniform (i.e the characteristic length scale of the cells vary across the domain), the analysis addresses the worst possible scenario and scales all cells by the largest for error computations. Thus, one would expect to see similar behaviour regarding numerical dissipation and dispersion.

\begin{table}[H]
\begin{center}
 \begin{tabular}{|c|c|c|}
   \hline
     Resolution &  $\delta \eta$ [m] &  $\delta t$ [s] \\
    \noalign{\hrule height 2pt}
    600 &  0.08 & 2.8 \\ \hline
    1000 &  0.43 & 2.8 \\ \hline
    2050 &  1.55 & 14.8 \\ \noalign{\hrule height 0.5mm}
    \end{tabular}
    \caption{Relative differences in wave height and arrival time of initial wave at gauge 3 between the coarser meshes and finest one (Fig \ref{fig:real_case}). $\delta \eta = $ the difference in maximum wave height and $\delta t = $ the time delay between the arrival of the maximum wave. }
    \label{tab:relative}
    \end{center}
\end{table}

\begin{figure}[H]
\begin{center}
  \includegraphics[width=\textwidth]{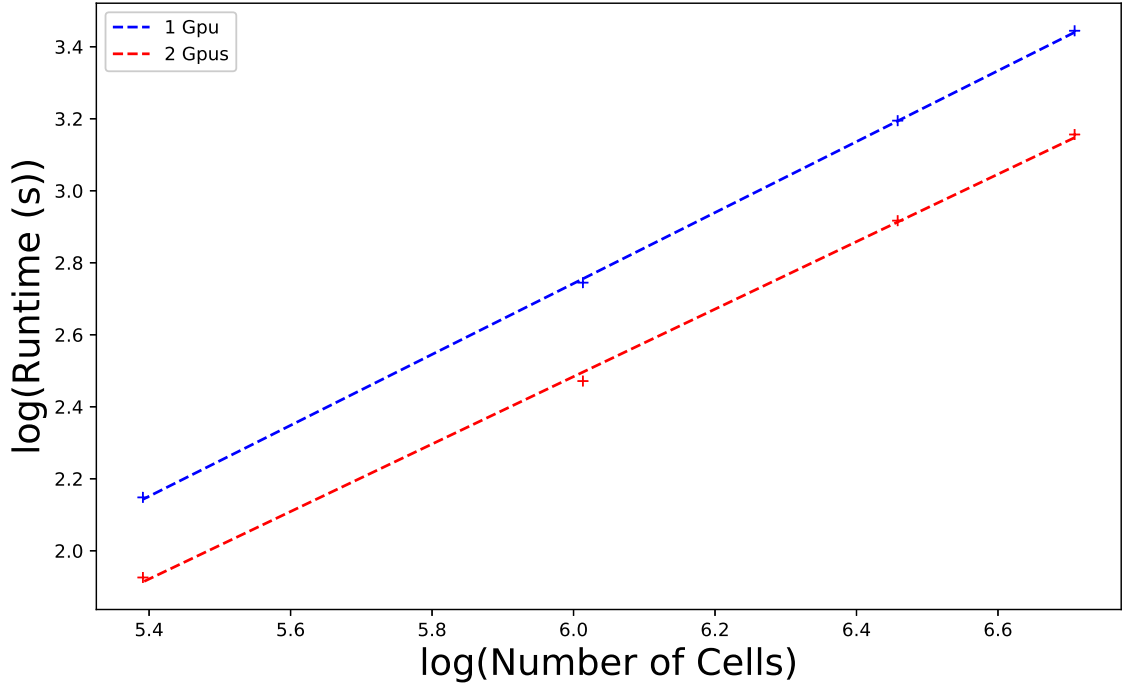}
  \caption{Runtimes of the various mesh resolutions with Volna-OP2 on a GPU cluster.}
  \label{fig:runtimes}
\end{center}
\end{figure}

Turning to the performance of Volna-OP2 on the GPU cluster, the table (\ref{tab:runtime}) and plot (Fig \ref{fig:runtimes}) summarise the runtimes for the various mesh resolutions.  One can see that we get a linear speed up of the runtimes. Those interested in the scalability of the code on other HPC architectures are referred to \cite{Reguly_Volna_2018}. This analysis of the relative errors  (\ref{tab:relative}) and computational efficiency informs the user on what resolution will provide an acceptable level of accuracy within a given time constraint.
\begin{table}[H]
\begin{center}
 \begin{tabular}{|c|c|c|c|}
   \hline
     Resolution, $\Delta x=\Delta y$ [m] & Number of cells  & Total runtime [s] - 1 GPU & Total runtime [s] - 2 GPUs \\
    \noalign{\hrule height 2pt}
    2050 &  246,187 & 140.7 & 84.3 \\ \hline
    1000 &  1,031,014 & 555.6 & 296.1  \\ \hline
    600 & 2,872,156 & 1565.8 & 826.0 \\ \hline
    450 & 5,109,001 & 2782.3 & 1433.1\\ \noalign{\hrule height 0.5mm}
    \end{tabular}
    \caption{Volna-OP2 performance on GPU cluster}
    \label{tab:runtime}
    \end{center}
\end{table}

\section{Concluding remarks}\label{Conclusions}

Based on the current study we can conclude that Volna-OP2 is a robust and efficient parallel solver for the NSWE. It is based on the finite volume scheme for spatial integration, implementing a MUSCL reconstruction and using the 2nd order Runge-Kutta scheme for integration in time. The scheme is conditionally stable with experimentally confirmed CFL=1.0. The code can handle complex geometries and simulate real-life cases.

The error analysis shows that it scales quadratically with refining the spatial mesh. However, reducing the time-step does not have a visible effect on the error. The solution amplitude decays in time, and one can observe both damping and phase shifts. So there is an energy leak from the system, which is expected due to the non-conservative Runge-Kutta scheme. However, the error can be minimised by reducing the spatial resolution.  The phase error is mostly negative, which corresponds to the phase lag and the artificial wiggles behind the wave front. However, it changes sign occasionally and this leads to the phase lead.

The real case simulations show the efficiency and scalability of the code run on a GPU cluster. The 1-hour realistic size tsunami model is simulated in $\sim$ 24 mins on the finest (450m resolution, $\sim 5.1M$ cells) mesh using two GPUs. However, the simulations have shown that comparable results can be achieved using a coarser mesh and reduced runtime. Therefore the users should be familiar with code pros and cons before setting up their simulation in order to get the physically meaningful and numerically accurate results. This acceptable level of accuracy and runtime trade off is an important decision when using Volna-OP2 to perform comprehensive sensitivity analysis tests and uncertainty quantification.

To conclude the paper we want to mention that Volna-OP2 is still under development. Therefore its analysis and benchmark testing play an important role for the code's continuous enhancement and improvement. The code is already an important tool for tsunami modelling community. However, we hope that our work will help to its wider adoption and will lead to discussion on the most suitable algorithms and software platforms for realistic tsunami modelling and prediction of its effects.

\section*{Acknowledgments}
The Irish Centre of High End Computing (ICHEC) is acknowledged for their computing resources. Daniel Giles is supported by a Government of Ireland Postgraduate Scholarship from the Irish Research Council (GOIPG/2017/68). Eugene Kashdan would like to acknowledge NVIDIA for the donation, within the framework of the Professor Partnership Programme, of the Tesla K40 GPU which was used for this research.  Serge Guillas acknowledges support from the Alan Turing Institute project ``Uncertainty Quantification of multi-scale and multiphysics computer models: applications to hazard and climate models'' as part of the grant EP/N510129/1 made to the Alan Turing Institute by EPSRC.

\end{document}